\documentclass[amsmath,amssymb,twocolumn]{revtex4}
\usepackage{amssymb}
\usepackage{graphicx}
\usepackage{verbatim}
\usepackage{dcolumn}
\usepackage{bm}

\begin{document}

\title{Green functions and correlation functions of a solvable
$S=1$ quantum Ising spin model with dimerization}

\author{ Zhi-Hua Yang$^{1}$, Li-Ping Yang$^2$, Hai-Na Wu$^{3}$,
Jianhui Dai$^{1}$, and Tao Xiang$^{4,2}$}

\address{$^1$Zhejiang Institute of Modern Physics, Zhejiang University,
Hangzhou 310027,  China\\
$^2$Institute of Theoretical Physics, Chinese Academy of
Science, P.O. Box 2735, Beijing 100080, China\\
$^3$College of Science, Northeastern University, Shengyang 110006,
China\\
$^4$Institute of Physics, Chinese Academy of Sciences, P.O. Box 603,
Beijing 100080, China }

\date{\today}

\begin{abstract}
This is a supplementary material of our recent paper\cite{yangPRB},
where a class of exactly solvable $S=1$ quantum Ising spin models
were studied based on the hole decomposition scheme. Here we provide
some details for the Green functions, the spin-spin correlation
functions, as well as the spin susceptibility in the presence of
dimerization.

\end{abstract}


\maketitle

\section{Introduction}
%In this supplemental material, we  will discuss the green functions
%and the spin-spin correlation at zero and finite temperature.

In Ref.\cite{yangPRB} we have studied a class of the $S=1$ spin
chains with the nearest neighbor Ising coupling and both transverse
and longitude single-ion anisotropy by a combinational use of a hole
decomposition scheme and a recursive method. These models include
the first example of the dimerized $S=1$ quantum spin chain where
all the eigen states can be solved exactly. In this supplementary
material we present some detailed derivations for the physical
quantities of the $S=1$ dimerized chain. All the notations are the
same as in Ref.\cite{yangPRB}. In Sec.~\ref{sec:green}, we discuss
the Green functions of the uniform or dimerized chains,
respectively. In Sec.~\ref{sec:correlation}, we study the
longitudinal spin-spin correlation function at zero- or
finite-temperatures. In Sec.~\ref{sec:appendM} and
\ref{sec:partition} we list some detailed formulae for the segmented
M-matrices and the partition functions.

\section{Green functions}\label{sec:green}

\subsection{Green functions of the uniform spin
segments}\label{app:realgreen} The original $S=1$ quantum Ising
model is mapped onto a large family of the segmented $S=1/2$
transverse Ising models classified by the total number of
holes\cite{Oitmaa, YangPRL}. These segmented $S=1/2$ models are then
solved by introducing the Bogoliubov fermionic quasi-particle
operators $\eta_k^{\dagger}$ and $\eta_k$ as defined in Eq.~(14) in
Ref.~\cite{yangPRB}. Inversely, we have
\begin{equation*}
\begin{split}
c_j^{\dagger}=\sum_k\frac{\Phi_{kj}+\Psi_{kj}}{2}\eta_k^{\dagger}
+\frac{\Phi_{kj}^*-\Psi_{kj}^*}{2}\eta_k,\\
c_j=\sum_k\frac{\Phi_{kj}^*+\Psi_{kj}^*}{2}\eta_k
+\frac{\Phi_{kj}-\Psi_{kj}}{2}\eta_k^{\dagger} .
\end{split}
\end{equation*}

The Green function, or the two-point correlation function, is
defined by
\begin{equation}
G_{jq}\equiv\langle F_j^{(-)}F_q^{(+)}\rangle,
\end{equation}
where $F_j^{(\pm)}\equiv c_j^\dagger\pm c_j$ .

For the uniform system, the wavefunctions $\Phi_{kj}$ and
$\Psi_{kj}$ can be taken as real, we have
\begin{equation}
\begin{split}\label{realfermion}
F_j^{(-)}=\sum_k\Psi_{kj}(\eta_k^{\dagger}-\eta_k)~,\\
F_j^{(+)}=\sum_k\Phi_{kj}(\eta_k^{\dagger}-\eta_k)~.
\end{split}
\end{equation}
The Green function can be then expressed as
\begin{eqnarray*}
G_{jq}(\beta)=-\sum_{k}\Psi_{kj}\Phi_{kq}\tanh[\beta\Lambda(k)/2]~.
\end{eqnarray*}
Note that $\tanh[\beta\Lambda(k)/2]\rightarrow 1$ at the ground
state ($\beta\rightarrow \infty$), so we have
\begin{eqnarray*}
 G_{jq}(\beta\rightarrow\infty)=-\sum_{k}\Psi_{kj}\Phi_{kq}~.
\end{eqnarray*}

We denote the wavefunctions for the chain with periodic boundary
condition (cyclic) and open boundary condition (free ends) by
($\Phi^c$, $\Psi^c$) and ($\Phi^f$, $\Psi^f$), respectively. Then we
have
\begin{eqnarray}\begin{split}\label{eq:phicpsic}
\Phi^c_{kj}&=\begin{cases}\sqrt{2/l}\sin jk~,~k>0,\\
\sqrt{2/l}\cos jk~,~k\leq0,\end{cases}\\
\Psi^c_{kj}&=-\frac{D}{\Lambda(k)}\left[(1+\lambda\cos
k)\Phi_{kj}^c+\lambda\sin k\Phi_{-kj}^c\right],\end{split}
\end{eqnarray}
where $l$ is the length of the segment. The Green function is
\begin{eqnarray}\label{eq:greenc}
&&G^c_{r}=L_r+\lambda L_{r+1},
\end{eqnarray}
where $r\equiv|j-q|$ and $L_{r}$ was defined in
Refs.~\cite{Lieb61,Pfeuty}
$$L_r=\frac{1}{\pi}\int_0^{\pi}
dk\frac{1}{\sqrt{1+\lambda^2+2\lambda\cos k}}\cos kr.$$

Similarly,
\begin{eqnarray}\begin{split}\label{eq:phif}
\Phi_{kj}^f&=A_k\sin(j-q+1)k,~\\
\Psi_{kj}^f&=A_k\delta_k\sin jk,~\end{split}
\end{eqnarray}
where
\begin{eqnarray}
A_k=\frac{1}{2}\left[2l+1-\frac{\sin{(2l+1)k}}{\sin
k}\right]^{-1/2}.
\end{eqnarray}
Consequently, we have
\begin{eqnarray}\label{eq:greenf}
G^f_{jq}=-\sum_kA_k^2\delta_k\sin jk\sin(j-q+1)k.
\end{eqnarray}

At the finite temperatures, we need to add the factor
$\tanh[\beta\Lambda(k)/2]$ to Eqs.(\ref{eq:greenc}) and
(\ref{eq:greenf}).

\subsection{Green functions of the dimerized
segments}\label{app:dimergreen}

In  the presence of dimerization, the wavefunctions $\Phi_{kj}$ and
$\Psi_{kj}$ are complex in general. So we now have,
\begin{equation}
\begin{split}\label{fermion}
F_j^{(-)}=\sum_k\Psi_{kj}\eta_k^{\dagger}-\Psi_{kj}^*\eta_k,\\
F_j^{(+)}=\sum_k\Phi_{kj}\eta_k^{\dagger}+\Phi_{kj}^*\eta_k.
\end{split}
\end{equation}
Then, the Green function is expressed by
\begin{equation}
\begin{split}\label{greentemperature}
G_{jq}=&\sum_{k}(\Psi_{kj}\Phi_{kq}^*+\Psi_{kj}^*\Phi_{kq})\langle
     \eta_k^\dagger\eta_k\rangle-\sum_k\Psi_{kj}\Phi_{kq}^*.
      \end{split}
      \end{equation}
Where,
$\langle\eta_k^\dagger\eta_k\rangle=[\exp{(\Lambda_k/(k_BT))}+1]^{-1}$,
satisfying Fermi-Dirac statistics. At the zero temperature, the
Green function can be written as
\begin{eqnarray}
\label{green} G_{jq}=D_jY[j,q]+2J_jY[j+1,q],
\end{eqnarray}
where
\begin{eqnarray}\label{eq:young}
Y[j,q]&=&-\sum_k\frac{e^{i(j-q)k}}{\Lambda(k)}[1+(-1)^{j+q}\gamma^*\gamma\nonumber\\
&&~~~~~~~+(-1)^j\gamma+(-1)^q\gamma^*].
\end{eqnarray}

The dimerization parameter $\gamma$ is defined by
 \begin{equation}
 \begin{split}
 \gamma=\frac{1-\tau}{1+\tau}
  \end{split}
 \end{equation}
with $\tau$ being determined by Eqs.~(19) in Ref.~\cite{yangPRB}.

Generally, $\tau$ has two solutions, corresponding to the
upper/lower signs of $\pm$  respectively in  Eqs.~(19) in
Ref.~\cite{yangPRB}. In order to numerically calculate the Green
function, we need to express $Y[j,q]$-function in terms of real
variables. We introduce $p_{1,2}$, $q_{1,2}$ to express complex
$\gamma$ as follows.
\begin{equation} \gamma_1=p_1+iq_1,~~~\gamma_{2}=p_{2}+iq_{2},
\end{equation} $p_{1,2}$ and $q_{1,2}$ are the real and imaginary
parts of $\gamma_{1,2}$, respectively,
\begin{eqnarray*}
p_{1,2}&=&\frac{b_1^2+b_2^2+4b_1b_2\cos2k-\left(\zeta_1\mp\zeta_2\right)^2}
{\left[(b_1+b_2)\cos
k-\zeta_1\pm
\zeta_2\right]^2+(b_2-b_1)^2\sin^2k},\\
q_{1,2}&=&\frac{-2(b_2-b_1)\sin k\left[(b_1+b_2)\cos
k+\zeta_1\mp\zeta_2\right]}{\left[(b_1+b_2)\cos k-\zeta_1\pm
\zeta_2\right]^2+(b_2-b_1)^2\sin^2k},
\end{eqnarray*}
where the subscript ${1}$  corresponds to the upper case, the
subscript $2$ corresponds to the lower case. $\zeta_{1,2}$ are given
by
\begin{eqnarray*}
\zeta_1&=&{(a_2-a_1)}/{2},\\
\zeta_2&=&\Gamma^2\sqrt{1-P+Q\cos2k}~.
\end{eqnarray*}
where $a_1$, $a_2$, $P$, $Q$ and $\Gamma$ are defined in
Ref.\cite{yangPRB}.

For convenience, we divide $k$-region $[-\pi, \pi)$ into two
subregions: ($I$) for $[-\pi/2,\pi/2)$ and ($II$) for
$[-\pi,-\pi/2)\cup[\pi/2,\pi)$, respectively. Thus $G_{jq}$ can be
expressed by
\begin{equation}
G_{jq}=G_{jq}^{(I)}+G_{jq}^{(II)}.
\end{equation}

In Region ($I$), because of the symmetry between $k$ and $-k$, the
Green function can be reduced in $(0,\pi/2)$,
\begin{equation}
\begin{split}
G_{jq}^{(I)}
=&-\sum_{(0,\pi/2)}\frac{2}{\Lambda_{-1}(k)}\{D_j[1+(-1)^{j+q}(p_1^2+q_1^2)\\
&+(-1)^jp_1+(-1)^qp_1]\cos(j-q)k\\
&+2J_j[1+(-1)^{j+q+1}(p_1^2+q_1^2)\\
&+(-1)^{j+1}p_1+(-1)^qp_1]\cos(j-q+1)k\}~.
\end{split}
\end{equation}
A similar Green function can be obtained for Region ($II$). The
function $Y[j,q]$ can be rewritten as
\begin{equation}
\begin{split}
Y[j,q]=&-\sum_{(0,\pi/2)}\frac{2}{\Lambda_{-1}(k)}
[1+(-1)^{j+q}(p_{1}^2+q_{1}^2)\\&+(-1)^jp_1+(-1)^qp_1]
\cos(j-q)k\\
&-\sum_{(\pi/2,\pi)}\frac{2}{\Lambda_{-2}(k)}[1+(-1)^{j+q}
(p_{2}^2+q_{2}^2)\\&+(-1)^jp_{2}+(-1)^qp_{2}]\cos(j-q)k. \label{yjl}
\end{split}
\end{equation}
So it is convenient to express the total Green function
Eq.~(\ref{green}) in terms of $Y[j,q]$. In the dimerization case,
there are four such Green functions associated with the four
different parity combinations of the segments.

\section{Correlation functions}
\label{sec:correlation}
\subsection{Zero temperature}
In this subsection, we discuss the spin-spin correlations at zero
temperature. In Ref.~\cite{yangPRB} we show that the ground state
has no hole if $D_z>-\Delta_h(0)$, otherwise, it has holes once
$D_z\le -\Delta_h(0)$. In the latter case, the holes break the
original chain into segments. We note that only the intra-segment
spin-spin correlations are non-zero.

For $D_z>-\Delta_h(0)$, the spin-spin correlation function of $S^z$
is defined by $C^z_{mn}=\langle\Psi_0|S_m^zS_n^z|\Psi_0\rangle$,
where $|\Psi_0\rangle$ is the normalized ground state of the
Hamiltonian. By use of the Jordan-Wigner transformation, one has
\begin{equation}
C_{mn}^z =\langle \Psi_0| F_m^{(-)}F_{m+1}^{(+)}F_{m+1}^{(-)}\cdots
F_{n-1}^{(-)}F_{n}^{(+)}|\Psi_0\rangle.
\end{equation}

It is straightforward to show that
$\langle\Psi_0|F_j^{(\pm)}F_q^{(\pm)}
|\Psi_0\rangle=\pm\delta_{jq}$. By further utilizing the Wick
Theorem, we find that
\begin{equation}\label{eq:rhoz}
C_{mn}^z= \left |
\begin{array}{cccc}
G_{m,m+1}     & G_{m,m+2}     & \cdots        & G_{m,n}  \\
G_{m+1,m+1}   & G_{m+1,m+2}   & \cdots       & G_{m+1,n}\\
\vdots        & \vdots        & \ddots           & \vdots   \\
G_{n-1,m+1}   & G_{n-1,m+2}   & \cdots      & G_{n-1,n}
\end{array}
\right |,
\end{equation}
for $n > m$, where, $G_{jq}=\langle\Psi_0|F_j^{(-)}F_q^{(+)}|\Psi_0
\rangle=-\langle\Psi_0|F_j^{(+)}F_q^{(-)}|\Psi_0 \rangle$.

The general expression of $G_{jq}$ is derived in
Sec.~\ref{app:realgreen} for the uniform chain and in
Sec.~\ref{app:dimergreen} for the dimerized chain respectively. In
general, one has
\begin{eqnarray}\label{eq:green1}
G_{jq}=D_jY[j,q]+2J_jY[j+1,q],
\end{eqnarray}
where $Y[j,q]$ is given by Eq.~(\ref{yjl}). For a uniform system,
$Y[j,q]=Y[q,j]=\frac{1}{D}L_{j-q}$.

\subsection{Finite temperatures}\label{sec:Temcorrelation}

At finite temperatures, the contribution from $p\neq0$-sector should
be taken into account. A recursion formula similar to Eq.~(36) in
Ref.~\cite{yangPRB} can be derived for the correlation function as
following
\begin{eqnarray}
\sum_{m,n}^LC_{mn}^z(\beta) &=& \frac{1}{Z(L)} \sum_{p=0}^{L}
\sum_{l=0}^{L-p} \sum_{m,n}^l  \alpha^p(p+1) \rho_{mn}^z \nonumber
\\ && z(l) Z^{(p-1)}(L-p-l).
\end{eqnarray}
Where, $\rho_{mn}^z$ is the correlation function of individual
segments. It has a similar form with that in Eq.~(\ref{eq:rhoz}),
but now $G_{jq}$ should be replaced by $G_{jq}(\beta)$.

\begin{figure}[h]
\includegraphics[width=0.95\columnwidth, bb=13 13 300 230]{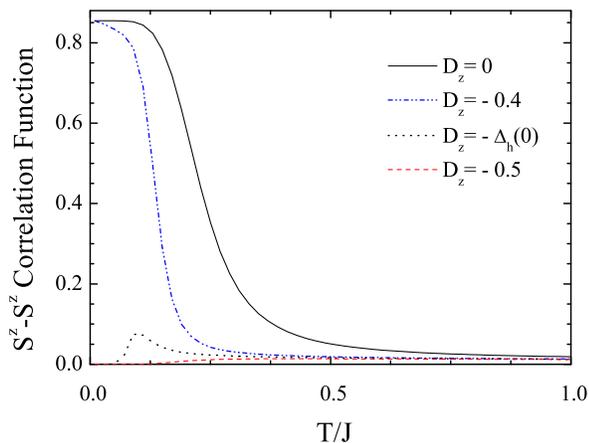}
\caption{Temperature dependence of the spin-spin correlation
function in a uniform spin chain with $\lambda=1.5$.}
\label{fig:correlationuniform}
\end{figure}

In Fig.~\ref{fig:correlationuniform}, we plotted the temperature
dependence of the spin-spin correlation function per site,
$\sum_{m,n}^LC_{mn}^z(\beta)/L$. We find that when
$D_z\leq-\Delta_h(0)$, the correlation function approaches to zero
in the limit $T\rightarrow0$. This indicates that the ground state
is in the hole condensation phase. On the other hand, when
$D_z>-\Delta_h(0)$, the correlation function approaches to a finite
value (about 0.85 for the two cases shown in the figure) in the zero
temperature limit.

\section{Spin susceptibility}

The spin susceptibility of the $S=1$ QIM can be also calculated
using the recursion formula introduced in the previous section. To
do this, one needs to first evaluate the partition functions of each
$S=1/2$ Ising segments in the applied magnetic field $\xi$, denoted
by $z(l_n, \xi)$. The partition function of the original $S=1$ QIM
is then given by $
Z(L,\xi)=\sum_{p=0}^L\sum_{\{l_n\}}\prod_{n=1}^{p+1}z(l_n,\xi)\alpha^p
$.  In terms of the segment magnetization
$m(l_n,T)=-\frac{1}{\beta}\frac{\partial \ln
z(l_n,\xi)}{\partial\xi}$ and the segment susceptibility
$\chi(l_n,T)=\frac{\partial m(l_n,T)}{\partial\xi}$, the total
susceptibility $\chi(T)$ at zero-magnetic field can be expressed as
\begin{eqnarray}
\chi(T)&=&\frac{1}{Z(L)}\sum_{p=0}^{L} \sum_{l=0}^{L-p}
\alpha^p(p+1) \nonumber \\
&& \chi(l,T)z(l)Z^{(p-1)}(L-p-l) .
\end{eqnarray}

Thus the hole decomposition scheme provide an alternative approach
to calculate the susceptibility of the $S=1$ QIM. This approach is
efficient provided that the susceptibilities of the corresponding
$S=1/2$ TIM's with varying chain length $L$ are available. We note
that the susceptibility of the $S=1/2$ TIM has already been studied
by a number of groups\cite{Pfeuty70,Ovchinnikov,skew}. So in
principle these results could be used in the numerical study of the
susceptibility of the $S=1$ QIM.

\section{Diagonalization of the M-matrix}\label{sec:appendM}

For a periodic spin chain, the diagonalization of the M-matrix has
been discussed in Sec.~IV~A in Ref.~\cite{yangPRB}. Here we consider
the diagonalization of this $l\times l$ M-matrix for an open spin
chain with the length $l$. The aim here is to solve the following
eigen equation
\begin{eqnarray}\label{appendM-Mphi}
M \Phi_k= \Lambda^2(k)\Phi_k
\end{eqnarray}
in various cases, where $\Phi_k(j)$'s take the form of Eqs.~(23) in
Ref.~\cite{yangPRB}.

We assume that the two ends of the open chain are located at the
sites $r_1$ and $r_2$, respectively. $r_1$ and $r_2$ can be either
odd or even, so there are four kinds of $M$-matrices. In the
following, we will present the results for each cases.

\subsection{$(r_1, r_2)=(odd,\, even)$}

In this case, the matrix $M$ is defined by
\begin{equation} \label{eq:M1}
M =
\begin{pmatrix}
a_0   & b_1   & 0    &\cdots          & 0      & 0\\
b_1   & a_2   & b_2   &\cdots           & 0      & 0\\
0     & b_2   & a_1   &\cdots         & 0      & 0\\
\cdots&\cdots &\cdots &\cdots   & \cdots &\cdots \\
0     & 0     & 0     &\cdots     &  a_1   & b_1   \\
0     & 0     & 0     &\cdots        & b_1    & a_2
\end{pmatrix},
\end{equation}
where $a_{1,2},~b_{1,2}$ are defined in the main text and
$a_0=D_1^2$.

The energy spectra can be solved following the approach introduced
in Section IV. The result is given by
\begin{eqnarray*}
\Lambda^2(k) & = &\frac{1}{e^{2ik}-t_ee^{-2ik}} [b_1\tau(e^{ik}-t_oe^{-ik})\\
&& + a_2(e^{2ik}-t_ee^{-2ik})+b_2\tau (e^{3ik}-t_oe^{-3ik}) ],
\end{eqnarray*}
The reflection parameters are
\begin{eqnarray}
t_o&=&e^{2i(l+1)k},~~\\
t_e&=&\frac{t_o(b_1e^{ik}+b_2e^{-ik})}{(b_1e^{-ik}+b_2e^{ik})}.\nonumber
\end{eqnarray}
Then, the secular equation is given by
\begin{eqnarray}\label{appendMoe-secular}
&&\left[(a_2-a_1)\pm W\right] [b_1\sin(l+2)k+b_2\sin lk] \nonumber \\
&= &\frac{2(a_0-a_1)(b_1^2+b_2^2+2b_1b_2\cos 2k)\sin lk}{b_2},
\end{eqnarray}
where $W$ is defined as in Eq.~(20) in Ref.~\cite{yangPRB}.

Other cases can be solved by the same way and the results are listed
below.

\subsection{$(r_1, r_2)=(odd,\, odd)$}

The reflection parameters $t_{o,e}$ are
\begin{eqnarray}
t_e&=&e^{2i(l+1)k},\\
t_o&=&\frac{t_e(b_1e^{-ik}+b_2e^{ik})}{(b_1e^{ik}+b_2e^{-ik})}.\nonumber
\end{eqnarray}
The secular equation is
\begin{eqnarray}\label{appendMoo-secular}
&&\left[(a_1-a_2)\pm
W\right]\left[b_1\sin(l-1)k+b_2\sin(l+1)k\right]\nonumber\\
&=&\frac{2b_2(b_1^2+b_2^2+2b_1b_2\cos2k)\sin(l+1)k}{a_0-a_1}.
%&&2b_2(b_1^2+b_2^2+2b_1b_2\cos(2k))\sin(l+1)k=\nonumber\\
%&&(a_0-a_1)\left[b_1\sin(l-1)k+b_2\sin(l+1)k\right]\left[(a_1-a_2)\pm
%W\right].
\end{eqnarray}

\subsection{$(r_1, r_2)=(even,\, even)$}

The reflection parameters $t_{o,e}$ are
\begin{eqnarray}\label{appendMee-tote}
t_e&=&e^{2i(l+1)k},\\
t_o&=&\frac{t_e(b_1e^{ik}+b_2e^{-ik})}{(b_1e^{-ik}+b_2e^{ik})}.\nonumber
\end{eqnarray}
The secular equation is
\begin{eqnarray}\label{appendMee-secular}
&&\left[(a_2-a_1)\pm
W\right]\left[b_1\sin(l+1)k+b_2\sin(l-1)k\right]\nonumber\\
&=&\frac{2b_1(b_1^2+b_2^2+2b_1b_2\cos2k)\sin(l+1)k}{a_3-a_2}
\end{eqnarray}
where, $a_3=D_2$.

\subsection{$(r_1, r_2)=(even,\, odd)$}

The reflection parameters $t_{o,e}$ are
\begin{eqnarray}
&&t_o=e^{2i(l+1)k},\\
&&t_e=\frac{t_o(b_1e^{-ik}+b_2e^{ik})}{(b_1e^{ik}+b_2e^{-ik})}.\nonumber
\end{eqnarray}
The secular equation is
\begin{eqnarray}\label{appendMeo-secular}
&&\left[(a_1-a_2)\pm
W\right]\nonumber\\&=&\frac{2b_1[b_1\sin(lk)+b_2\sin(l+2)k]}{a_3-a_2}.
\end{eqnarray}

\section{The partition functions of segments}\label{sec:partition}
The partition function of individual segment of length $l$ and
parity $(r_1, r_2)$ (defined in Sec.~\ref{sec:appendM}) is given by
\begin{equation}
z_{(r_{1}, r_{2})}(l)=\prod_{\substack{k_1\in (0,\pi/2),\\
k_2\in
(\pi/2,\pi)}}\cosh\left[\frac{\beta\Lambda_{1}(k_1)}{2}\right]\cosh\left[\frac{\beta\Lambda_{2}(k_2)}{2}\right],
\end{equation}
where, $k_{1,2}$ satisfy the corresponding secular equations.

\section*{Acknowledgments}
This work was supported in part by the National Natural Science
Foundation of China, the national program for basic research of
China (the 973 program), the PCSIRT (IRT-0754), and SRFDP
(No.J20050335118) of Education Ministry of China.

\end{document}